\documentclass[acmlarge,authorversion,nonacm]{acmart}
\usepackage[utf8]{inputenc}%
\usepackage{tabularx}%
\usepackage{circledtext}%
\usepackage{setspace}%
\usepackage{color}%
\newcommand\todo[1]{{\color{black}{#1}}}%
\newcommand\ok[1]{{\color{black}{#1}}}%
\usepackage{caption}%
\usepackage{subcaption}%
\usepackage[T1]{fontenc}%
\copyrightyear{2025}
\acmYear{2025}
\renewcommand\footnotetextcopyrightpermission[1]{} 
\setcopyright{none}
\usepackage[T1]{fontenc}
\usepackage[utf8]{inputenc}
\usepackage[greek,english]{babel}

\begin{document}%

\title[DangerMaps]{%
DangerMaps: Personalized Safety 
Advice for Travel in Urban Environments using a
Retrieval-Augmented
Language Model
}%

\author{Jonas Oppenlaender}
\email{jonas.oppenlaender@oulu.fi}
\affiliation{%
  \institution{University of Oulu}
  \city{Oulu}%
  \country{Finland}%
}



\begin{abstract}%
Planning a trip into a potentially unsafe area is a difficult task.
We conducted a formative study on travelers' information needs, finding that most of them turn to search engines for trip planning.
Search engines, however, fail to provide easily interpretable results adapted to the context and personal information needs of a traveler.
Large language models (LLMs) create new possibilities for 
providing personalized travel safety 
advice.
To explore this idea, we developed \textit{DangerMaps}, a mapping system that assists its users in researching the safety 
of an urban travel destination, whether it is pre-travel or on-location.
DangerMaps plots 
safety 
ratings onto a map and provides explanations on demand.
%
This late breaking work specifically emphasizes the challenges of designing real-world applications with large language models.
We provide a detailed description of our approach to prompt design and highlight future areas of research.
\end{abstract}%




\begin{teaserfigure}
\centering%
 \begin{subfigure}[b]{0.19\linewidth}%
     \centering
     \includegraphics[width=\textwidth]{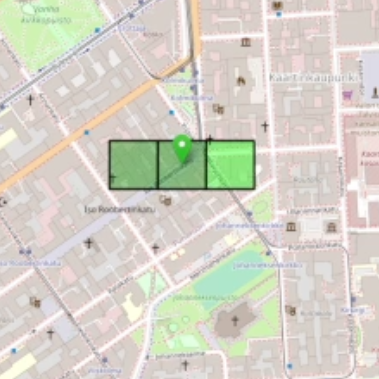}
     \caption{}
     \label{fig:teaser:a}
 \end{subfigure}
 \hfill
 \begin{subfigure}[b]{0.19\linewidth}
     \centering
     \includegraphics[width=\textwidth]{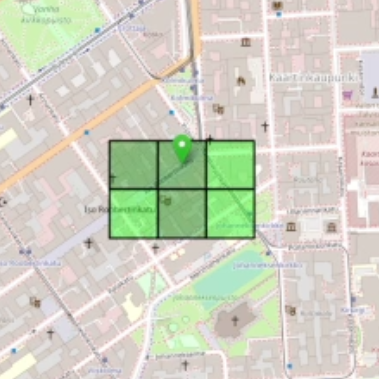}
     \caption{}
     \label{fig:teaser:b}
 \end{subfigure}
 \hfill
 \begin{subfigure}[b]{0.19\linewidth}
     \centering
     \includegraphics[width=\textwidth]{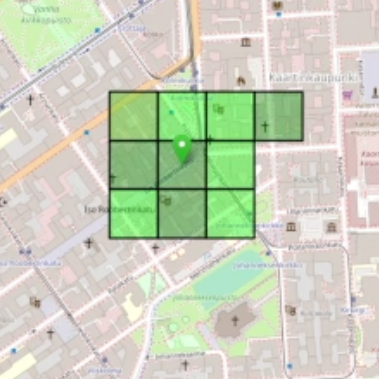}
     \caption{}
     \label{fig:teaser:c}
 \end{subfigure}%
 \hfill
 \begin{subfigure}[b]{0.19\linewidth}
     \centering
     \includegraphics[width=\textwidth]{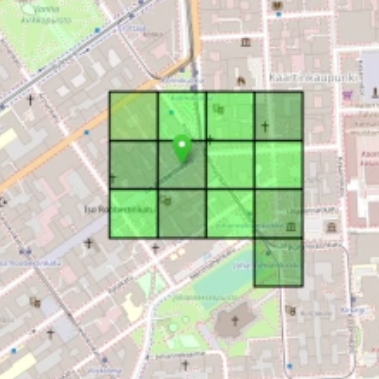}
     \caption{}
     \label{fig:teaser:d}
 \end{subfigure}%
 \hfill
 \begin{subfigure}[b]{0.19\linewidth}
     \centering
     \includegraphics[width=\textwidth]{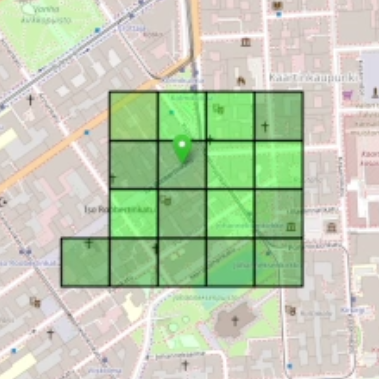}
     \caption{}
     \label{fig:teaser:e}
 \end{subfigure}%
\caption{
DangerMaps uses retrieval-augmented generation \cite{RAG} to plot personalized safety advice on a map.
Each map square of 75m side-length is color-coded according to a numeric safety rating from unsafe (0 -- red) to safe (100 -- green).
This figure demonstrates how safety ratings are queried and plotted in a spiral pattern, starting from the user's current location (or from a location that is being researched remotely).
The depicted location is in Helsinki, Finland, and safety ratings are personalized for a 35 year-old man traveling on foot and alone.
(Map data copyrighted by OpenStreetMap contributors, 2025).%
}%
\Description{A demonstration showcasing how DangerMaps maps safety ratings in a spiral pattern, starting near the current location.}%
\label{fig:spiral}%
\label{fig:teaser}%
\end{teaserfigure}%

\maketitle


\section{Introduction}%
\label{sec:introduction}%

\ok{%
Planning a trip can be a complex task, especially for trips to potential unsafe locations.
    For a given travel destination, safety-related questions might include:
    \textit{``Where is it safe for me to stay?''},
    \textit{``Which areas should I avoid at night?''},
    or in safety-critical situations,
    \textit{``Which direction should I head toward to get to safety?''}
When planning a trip---or when seeking information just-in-time in a location---numerous online information sources would need to be searched and reviewed to determine the answer to these questions.
%
%
Compounding this problem is that each traveler's perception of danger and safety is different.
    What is perceived as a risk (or potential danger) for one person in one location may not be perceived as a risk by another person in the same location.
Personal demographic attributes of a person (e.g., age and gender) determine the person's safety-related information needs.%
}%

\ok{%
Search engines fail to satisfy these personal information needs, because they may not personalize their results to safety-related contextual information.
Further complicating the search for safety-related travel information is that this information is dispersed on the Web and 
may be hard to find.
Therefore, researching the safety of a location is a sensemaking activity that requires high cognitive effort \cite{Sense_Making_206_Camera_Ready_Paper.pdf,informationforaging}. This type of cognitive activity involves foraging for information \cite{informationforaging}: tedious online research on neighborhoods, hotels, modes of travel, locations, and many other safety-related factors.
This highlights the need for solutions that provide \textit{personalized safety advice} to tourists and travelers.
}%



\ok{%
Large language models (LLMs) could provide an opportunity to support tourists and travelers in researching safety-related information for their travel destinations.
But while LLMs are potentially capable to adapt and personalize their responses to users' personal information needs, their intrinsic knowledge is not granular enough for detailed travel advice.
However, the LLMs' ability to understand instructions and learn in-context \cite{2005.14165.pdf,NEURIPS2022_8bb0d291,2204.02329.pdf,2021.naacl-main.185.pdf} could be employed to provide the LLM with the missing information needed to provide personalized safety advice to travelers.
    Given personal details about a traveler, such as basic demographics and whether the person is traveling alone or in a group, an LLM could provide a personal safety assessment
    for a given geographic location.
    For instance, the LLM could be provided with the following information: \textit{``I am a woman, 21 years of age, I travel on foot and alone, this is my current location, these are points of interest (POIs) in the vicinity.''}
    Given this information, the LLM could answer the safety-critical question: \textit{``How safe am I in this location?''} 
}%


\ok{%
We developed \textbf{DangerMaps}, an interactive mapping system using a large language model for the knowledge-intensive task of providing \textbf{personalized safety advice} to travelers.
This is one of the first applications harnessing the LLM's parametric world knowledge \cite{1818_emergent_world_representations.pdf}, retrieval-augmented generation \cite{RAG}, and in-context learning \cite{2005.14165.pdf,NEURIPS2022_8bb0d291,2204.02329.pdf,2021.naacl-main.185.pdf} for reasoning about the safety of a given location.
DangerMaps 
employs a prompt augmented with geolocation information, travel safety information, and geospatial points of interest (POIs), sourced from multiple websites.
This information is provided to the LLM which calculates a numeric safety rating for the specified location and its surroundings. This safety rating is personalized to the user, based on the user's age, gender, and mode of travel (solo versus group travel).
}%






\ok{%
In this paper, we demonstrate the feasibility of providing personalized safety advice for a given location with a large language model, and we describe the technical implementation of the DangerMaps prototype.
Our key focus is on highlighting the challenges of developing retrieval-augmented applications grounded in real-world context.
We describe our approach to designing the prompt for this system and highlight critical areas for future work and self-critically question the limitations to our approach.
%
%
%
In summary, we make the following contributions:%
\begin{itemize}%
    \item We demonstrate the feasibility of generating personalized travel safety advice with a retrieval-augmented large language model and implement this idea in DangerMaps, a system that plots personalized safety ratings onto a map. 
    \item  We 
        describe our prompt design process, which may serve as reference for other implementations.
    \item We 
        discuss open challenges and limitations of using LLMs for personalized safety advice, with general implications for the development of LLM-based applications.%
\end{itemize}%
}%


\section{Related Work}%
\label{sec:relatedwork}%

\subsection{Safety in Urban Environments}%
\label{sec:xxx}%
\ok{%

Urban environments present unique challenges to the safety of tourists.
One dimension is the fear of crime and the perceived risk of victimization. \citeauthor{2851581.2892400.pdf} addressed these aspects, identifying the ways in which fears affect not only individual emotions but also influence patterns of movement and activities within cities \cite{2851581.2892400.pdf}.
Fear can be more pronounced among certain groups within the population who may be more vulnerable than others.
\citeauthor{1952222.1952308.pdf} have focused on the fear and danger associated with nighttime in urban environments \cite{1952222.1952308.pdf,1979742.1979630.pdf}, shedding light on how perceived risks can alter behaviors and the sense of security within urban areas.

The literature has also pinpointed concerns related to gender and social status. For instance, \citeauthor{3025453.3025532.pdf} have emphasized the issue of women's safety in public spaces, investigating panic buttons as a way to enhance security \cite{3025453.3025532.pdf}. The subject of women's safe commutes in slums has been another area of research~\cite{2371664.2371675.pdf}.
\citeauthor{1978942.1979191.pdf} explored the unique safety considerations of homeless young people \cite{1978942.1979191.pdf}, while \citeauthor{3301019.3323898.pdf} have innovated in designing technology to support the safety of transgender women and non-binary people of color \cite{3301019.3323898.pdf}.
These works underscore the broader context of urban safety, recognizing the variations in risk perception and exposure based on gender, socioeconomic status, or membership in a vulnerable group.
These works contribute to a broader understanding of the diverse needs and challenges that different groups face in urban environments.


Managing risk in urban areas is more than a theoretical concern --- it represents an essential user need~\cite{1952222.1952308.pdf}.
The multifaceted nature of safety in urban environments requires a nuanced approach that recognizes the diversity of travelers' needs, experiences, and vulnerabilities. This leads us to an exploration of the challenges associated with information seeking for safe travel.%
}%
%
%
%
%
%
%
%
%
%
%
%
\subsection{The Challenges of Information Seeking for Safe Travel}%
\label{sec:related:infoseeking}%
\ok{%
%
Information search is an important aspect of 
planning a safe trip.
%
In this task, information has to be foraged and synthesized from disparate information sources \cite{informationforaging,3613905.3636322.pdf}, such as
    reviews on hotel booking websites,
    travel websites (e.g., Tripadvisor),
    and social media (Youtube, TikTok, 
    etc.).
%
Perhaps a more convenient way of finding safety-related information in a travel destination are the travel advisories maintained by governments for their citizens.
However, this information is often not very granular (i.e., not available for a given neighborhood or street, for instance).
Other sources, such as crime databases and crime maps maintained by the Police in some countries \cite{2851581.2892400.pdf}, may 
be outdated \cite{3491102.3501889.pdf} or simply not well suited for consumption and trip planning (e.g., Excel sheets with crime statistics).
These sources may also not be available for every city and every country, or they may be biased \cite{v1_stamped.pdf,2487788.2488075.pdf}.
The lack of recency and granularity of many of these offical sources make them impractical for researching the current safety of a specific location within a city.
Synthesis of these information sources for travel planning is a cognitively demanding task \cite{2308.07517.pdf}.
}%


\ok{%
Further complicating the search for information is that the search for travel information depends not only on many contextual factors (e.g., the travel destination), but also on the searcher~\cite{2818869.2818905.pdf}.
\citeauthor{2818869.2818905.pdf} investigated strategies for tourism information seeking
and found differences in search strategies between different user demographics \cite{2818869.2818905.pdf}.
This is not surprising, given that different users may have different information needs and information seeking habits.
    For instance, essential workers during the COVID-19 pandemic had a specific set of safety-related information needs~\cite{3555197.pdf}.
    Another example are young people who often turn to social media to satisfy their information needs \cite{doi:10.1177/0961000616631612}.
}%

\subsection{Language Models}%
\label{sec:related:llms}%
\ok{%
%
Pretrained language models show impressive versatility and performance on a large variety of downstream tasks, creating new possibilities for decision support in real-world applications.
%
One ongoing area of research is the world knowledge encoded in LLMs.
LLMs have been demonstrated to build emergent world representations \cite{1818_emergent_world_representations.pdf}.
Increasingly, LLMs are also being trained on multi-modal data, which can be expected to improve their world building abilities.
Research on real-world applications seeking to harness LLMs' world knowledge, however, are still an under-explored area of research.
The LLM's parametric knowledge combined with their ability to follow instructions would make them good candidates for providing personalized safety advice to travelers.
}%



%

\ok{%
However, since LLMs are very costly to train, they are updated infrequently.
The language model is frozen in time, and it will not be able to provide advice on current events.
To alleviate this problem, the LLM's ability to learn in-context can be leveraged to provide the LLM with up-to-date information on current events in the prompt.
This is referred to as retrieval-augmented in-context learning \cite{63c6c20dec4479564db21819_NEW_In_Context_Retrieval_Augmented_Language_Models.pdf}.
DangerMaps employs retrieval-augmented generation for providing personalized safety advice to travelers.
In the following section, we describe a formative study on people's safety-related information needs in urban environments.
}%






\section{Formative Study}%
\label{sec:pre-study}%
\ok{%
To inform the design of DangerMaps, we launched an online survey on Prolific, an online recruitment platform for academic surveys, exploring 
how people research the safety of a foreign city
and what people worry about when traveling.
    Compared to other popular crowdsourcing platforms, participants on Prolific are attentive, follow instructions, and provide meaningful answers \cite{journal.pone.0279720.pdf}.
The formative study informed a set of design requirements for our prototype. 
}%

\subsection{Survey Design and Participants}%
\label{sec:pre-survey}%
\ok{%
The short survey included six questions (including three demographic items) and focused on personal safety requirements of tourists in a foreign urban environment.
Participants ($N=26$) were asked to rate the importance of the concept of `safety' when they plan a trip into a foreign city, on a 5-point Likert scale (from 1 -- Not Important At All to 5 -- Extremely Important).
This was followed by two open-ended questions:
\begin{itemize}
    \item[Q1:] \textit{When you travel to a new city in a foreign country, what safety-related factors do you worry about the most?}
    \item[Q2:] \textit{As a tourist, how do you research the safety of a specific location in the city? And what information sources do you turn to?} Please list specific websites, if you know them.
\end{itemize}
The Prolific task was completed in 4~minutes and 51~seconds on average and each participant was paid {\pounds}0.60.
}%

\ok{%
The gender-balanced sample of 26~participants (13 men, 13 women, no non-binary) were between 20 and 49 years old ($M=29.3$ years) and well educated, with completed Bachelor degrees ($n=9$) and Master degrees ($n=9$) being most common.
The sample consisted of participants from
South Africa ($n=6$),
Portugal	($n=5$),
United Kingdom	($n=3$),
Greece	($n=2$),
Italy	($n=2$),
and other countries ($n=8$).
}%

\subsection{Survey Results}
\ok{%
To most participants, safety was extremely important. A clear majority of participants ($n=17$; 65.4\%) rated safety `extremely important' 
when planning a trip to a foreign city. Seven participants (26.9\%) responded with 4 on the given 5-point Likert scale, two with 3, and no participants thought safety was unimportant.
}%

\ok{%
We qualitatively analyzed the open-ended responses. This analysis was straightforward and did not require inter-rater agreement~\cite{McDonald_Reliability_CSCW19.pdf}.
The analysis revealed that the safety-related factors considered by participants are quite diverse.
with worries about robbery and theft being the most mentioned ($n=11$), followed by general crime (not specified in detail; $n=8$) and violence/aggression ($n=6$).
}%


\ok{%
We note that there was low agreement among participants on the sources used to research travel destinations.
About half of the participants mentioned they would complete their research on a major online search engine ($n=13$).
Nine participants mentioned different social media sites, chief among them Youtube ($n=3$).
Specialized travel advice websites, such as Lonelyplanet ($n=1$) and Tripadvisor ($n=4$), were mentioned by a surprisingly low number of participants.
}%


\ok{%
Overall, our survey highlights the diversity of the participants needs in researching their travel destinations.
Only one person mentioned online map services as data source, which may be indicative of a gap in the availability of Geographic Information Systems (GIS) supporting safe travel.
The survey also highlights the difficulty of finding safety-related information online, with many participants not having a go-to location on the web to look up the safety of a specific location.
Rather, almost all participants mentioned having to peruse more than one information source for researching the safety of the travel destination.
The granularity with which this information may be required (e.g., the safety of a particular hotel or neighborhood) drives participants to websites which may contain biased information (e.g., hotel review websites and social media).
This highlights the need for an unbiased, easily comprehensible solution that provides granular and personalized safety advice to tourists and travelers.
}%

\subsection{Design Goals}%
\label{sec:requirements}%
\ok{%
Based on the results of our formative study and the review of related literature, we formulate a set of requirements for a solution that provides personalized safety advice to travelers (see \autoref{tab:requirements}).
    While the functional requirements F1--F5 focus on the core functionalities essential for providing tailored safety advice, the non-functional requirements N1--N3 relate to quality and performance attributes that enhance user experience.
}%

\begin{table*}[!htb]%
\caption{Functional 
and non-functional 
requirements for our solution providing personalized travel safety advice.}
\label{tab:requirements}
\small
\begin{tabularx}{\textwidth}{llX}
\toprule
No. & Requirement & 
    Description \\
\midrule
F1 & Safety information & 
    Safety information should be pulled and evaluated from different sources to provide an assessment of a given location's safety. 
\\
F2 & Map visualization & 
    Safety recommendations should be plotted on a map, centered around a given location.
\\
F3 & Personalization & 
    The safety recommendations must be personalized and tailored to the traveler's background and current context to account for demographic diversity and individual travel preferences.
\\
F4 & Priority & 
    Since the traveler may already be located in a dangerous area, it is critical to give advice for this location first. The safety recommendations for the traveler's current location must take precedence over nearby recommendations.
\\
F5 & Explanations & 
    The service should provide explanations of its safety ratings.
\\
N1 & Accuracy & 
    The safety recommendations should reflect an honest estimate of the traveler's personal safety (or conversely, danger) in the specific location, considering all given information.
\\
N2 & Ease of understanding & 
    Safety recommendations must be quickly comprehensible. One look must give the traveler an understanding of whether a specific location is safe for the traveler or not, without having to sift through numerous online information sources.
\\
N3 & Low latency & 
    Safety recommendations need to be generated on-the-fly. As the traveler moves from location to location, a new set of safety recommendations must be generated with low latency.
\\
\bottomrule%
\end{tabularx}%
\end{table*}%

\ok{%
With the requirements in place, we proceed to describe the design of our solution
in the following section.
}%

\section{DangerMaps}%
\label{sec:method}%
\label{sec:dangermaps}%
\ok{%
This section presents the functionality of DangerMaps, how it addresses the design goals, our prompt design process, and the technical implementation.
}%

\subsection{Overview of Functionality}%
\label{sec:app-design}%
\ok{%
DangerMaps is a web application that plots personalized safety ratings on an interactive map (fulfilling design requirement F2).
The user interface (UI) is intentionally kept minimal in the spirit of a minimum viable product (MVP). The UI consists of a full-screen map onto which safety ratings are plotted in squares. 
The map can be zoomed and panned, and the user's current location is highlighted with a marker (F4, N2).
The ratings are color-coded, from red (0 -- unsafe) to safe (100 -- green), and therefore easy to understand and actionable (N2).
Each square in the map can be clicked to open a pop-up with explanations
    (F5).
Priority is given to the initial set of coordinates from which the safety mapping expands in a spiral pattern (F4, N3).
The safety ratings are personalized for the user's demographics (F3). For this feasibility study, we consider three variables:
    age, gender, and travel mode,
    indicating whether the user travels alone or in a group.
}

\begin{figure*}[htb]%
     \centering
     \begin{subfigure}[b]{0.22\textwidth}
         \centering
         \includegraphics[width=\textwidth]{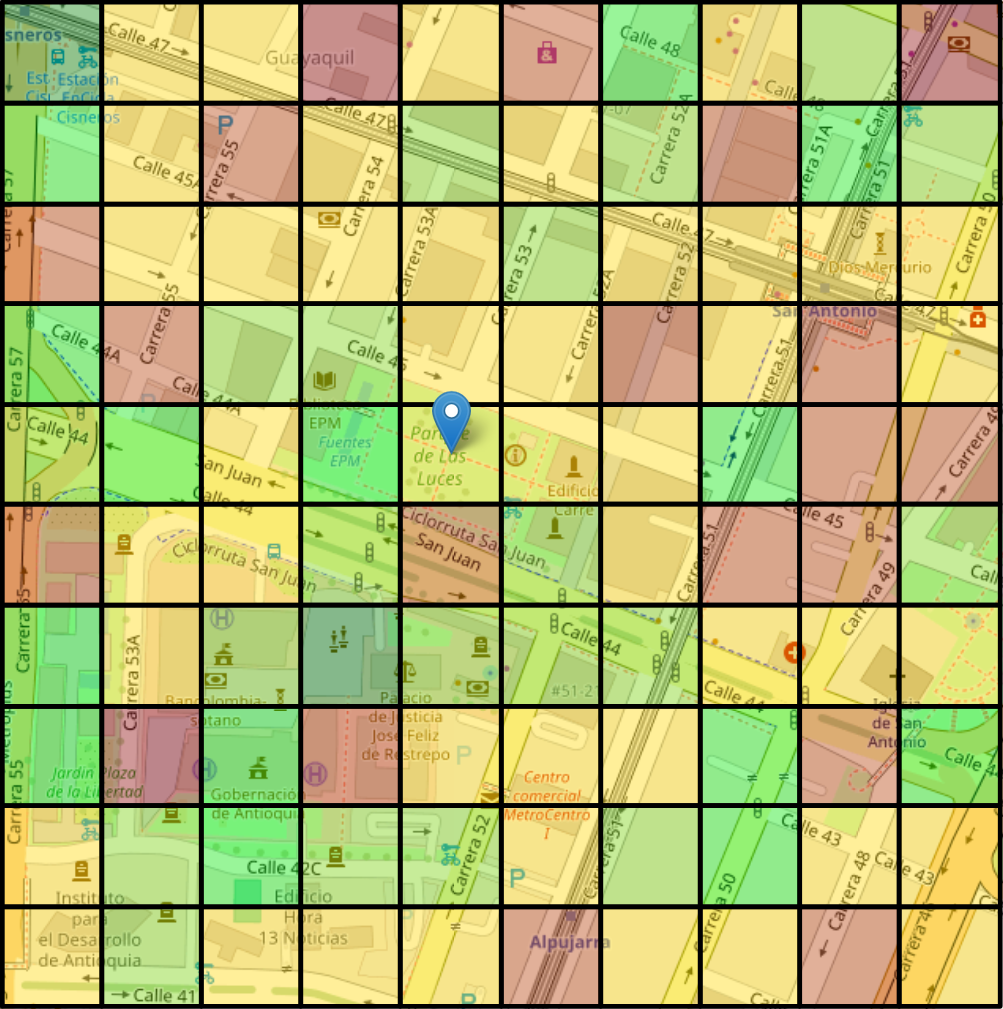}
         \caption{Medellin}
         \label{fig:dangermap:athens}
     \end{subfigure}
     ~~
     \begin{subfigure}[b]{0.22\textwidth}
         \centering
         \includegraphics[width=\textwidth]{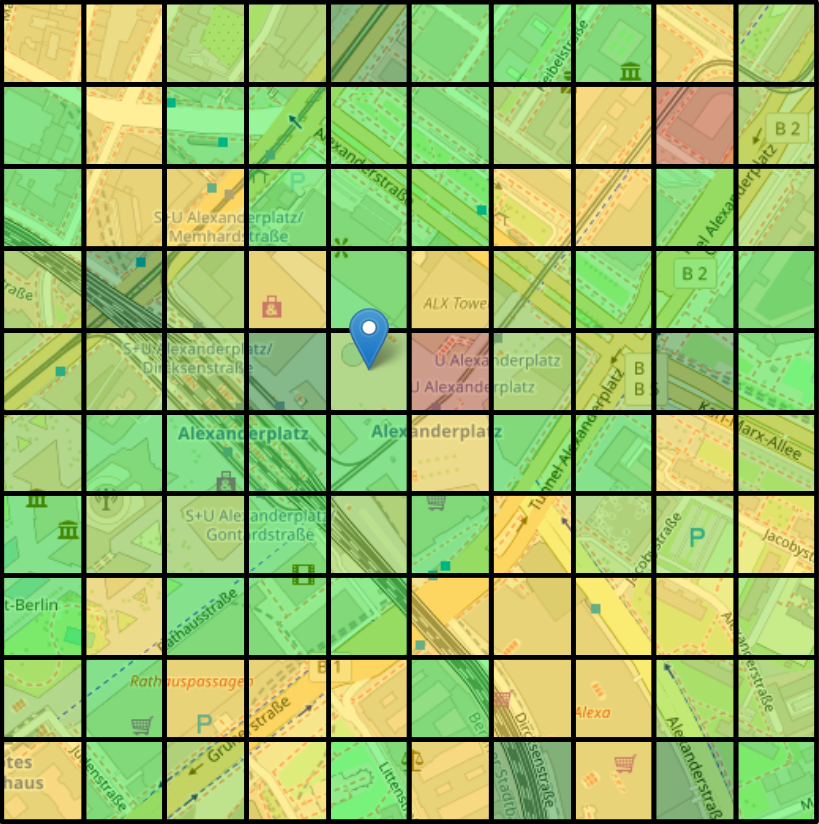}
         \caption{Berlin}
         \label{fig:dangermap:berlin}
     \end{subfigure}
     ~~
     \begin{subfigure}[b]{0.22\textwidth}
         \centering
         \includegraphics[width=\textwidth]{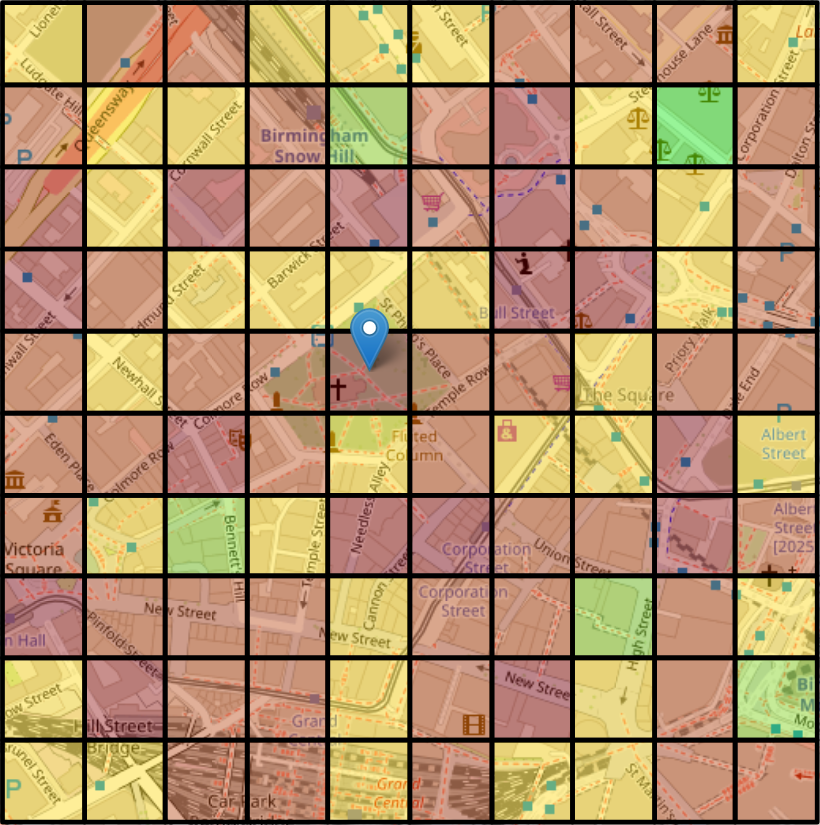}
         \caption{Birmingham}
         \label{fig:dangermap:birmingham}
     \end{subfigure}
     ~~
     \begin{subfigure}[b]{0.22\textwidth}
         \centering
         \includegraphics[width=\textwidth]{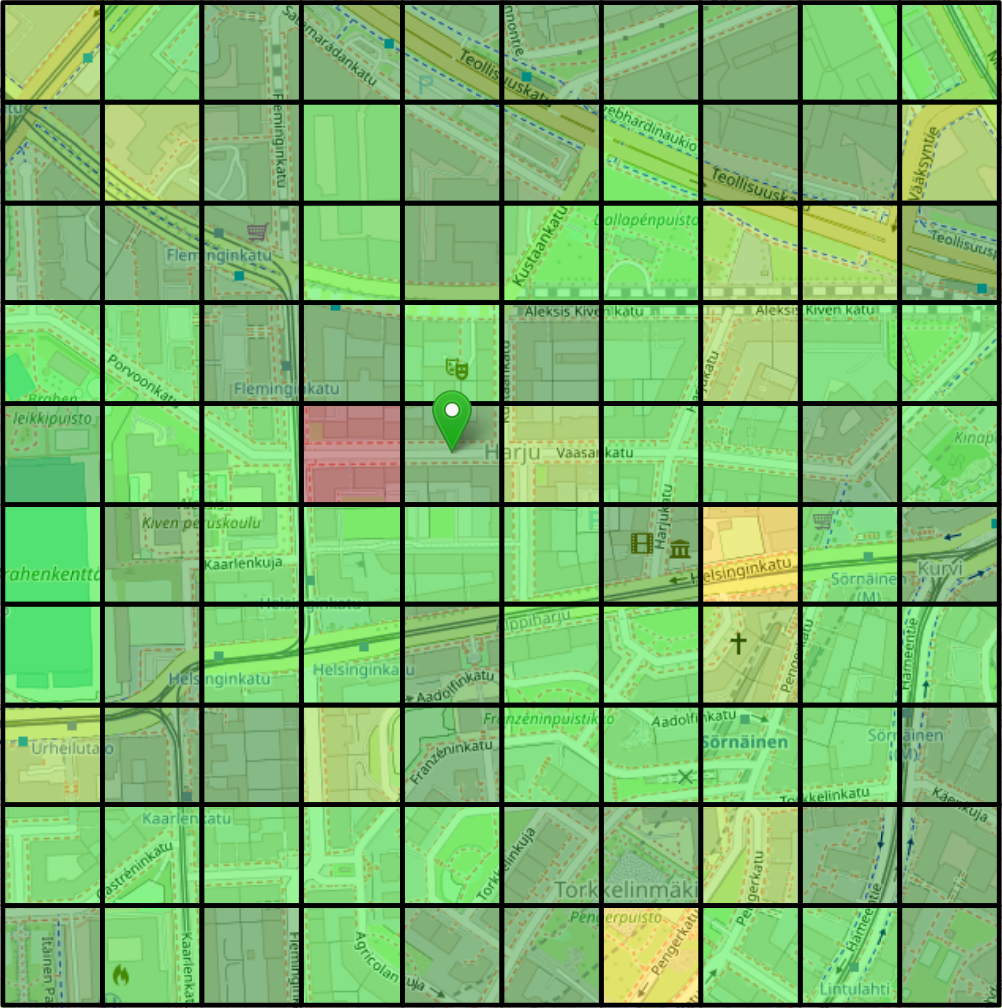}
         \caption{Helsinki}
         \label{fig:dangermap:helsinki}
     \end{subfigure}
\caption{Excerpts of DangerMaps for different international cities.
All plots were created for a \todo{solo-traveling man of 30~years of age at 12AM (midnight).}
    (Map data copyrighted OpenStreetMap contributors, 2025).}%
\label{fig:dangermaps}%
\end{figure*}

\begin{figure*}[!htb]%
\centering%
 \begin{subfigure}[b]{0.16\textwidth}%
\centering%
     \includegraphics[width=\textwidth]{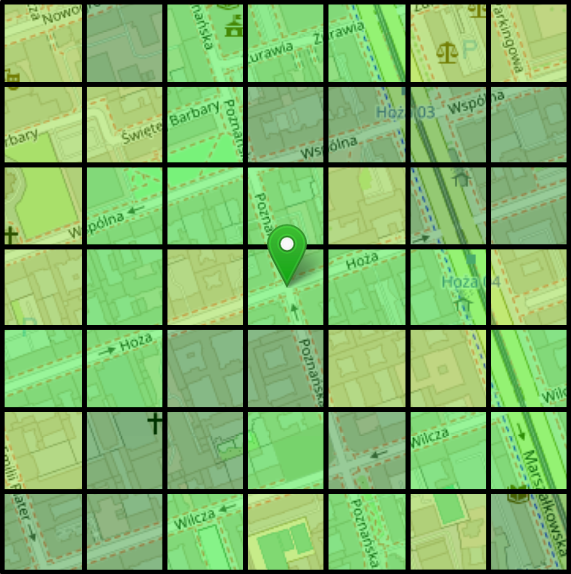}
     \caption{Woman\\~}
    \label{fig:personacomparison:a}
 \end{subfigure}
~~
 \begin{subfigure}[b]{0.16\textwidth}
\centering%
     \includegraphics[width=\textwidth]{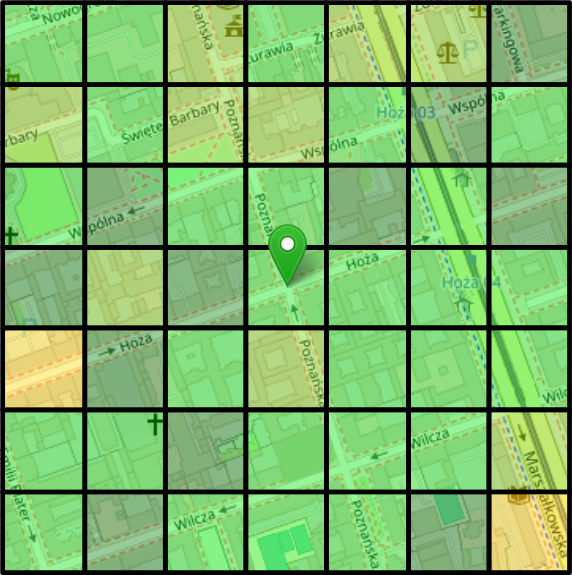}
     \caption{Transgender woman}
    \label{fig:personacomparison:b}
 \end{subfigure}
 ~~
 \begin{subfigure}[b]{0.16\textwidth}
\centering%
     \includegraphics[width=\textwidth]{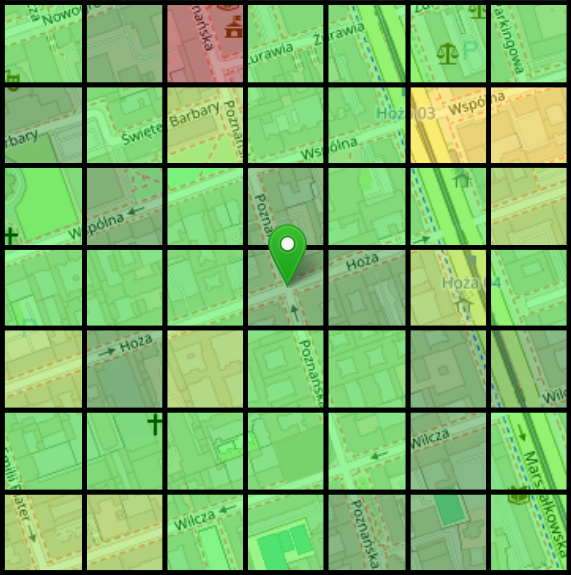}
     \caption{Man\\~}
    \label{fig:personacomparison:c}
 \end{subfigure}
 ~~
 \begin{subfigure}[b]{0.16\textwidth}
\centering%
     \includegraphics[width=\textwidth]{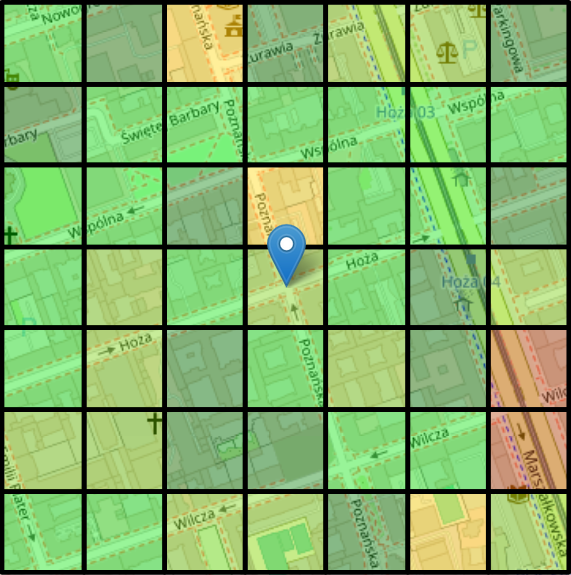}
     \caption{Blind man with a cane}
    \label{fig:personacomparison:c}
 \end{subfigure}
 ~~
 \begin{subfigure}[b]{0.16\textwidth}
\centering%
     \includegraphics[width=\textwidth]{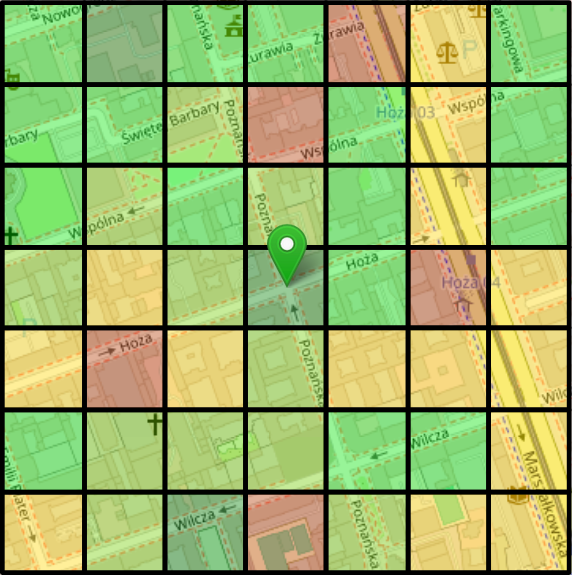}
     \caption{Homeless man\\~}
    \label{fig:personacomparison:c}
 \end{subfigure}
\caption{%
Personalized danger maps for five solo-traveling 23-year old personas
at a random location in Warsaw, Poland, 11AM.
\\
(Map data copyrighted OpenStreetMap contributors, 2025).
}%
\label{fig:personacomparison}%
\end{figure*}

We demonstrate the capabilities of DangerMaps in a series of screenshots.
%
\autoref{fig:teaser} depicts the time progression of plotted ratings.
%
%
\autoref{fig:dangermaps} compares safety ratings plotted for four different cities.
%
%
In \autoref{fig:personacomparison}, one can observe differences in safety ratings for five different personas.
\subsection{Technical Implementation}%
\label{sec:implementation}%

\ok{%
DangerMaps is implemented as a light-weight research prototype using standard web development technologies.
Leaflet\footnote{https://leafletjs.com}, a JavaScript mapping library,
and data from OpenStreetMap (OSM)\footnote{https://www.openstreetmap.org} are used for creating the interactive map.
Geospatial points of interest (POIs) are queried from OpenStreetMap's Overpass API.\footnote{https://wiki.openstreetmap.org/wiki/Overpass\_API}
OpenStreetMap's Nominatim API\footnote{https://nominatim.org} is used for performing a reverse geolocation lookup which takes a set of coordinates and returns the administrative area, city, neighborhood, and street.
For each map square, requests to the Overpass API, Nominatim API, and OpenAI's API are being made. The latter request cannot be parallelized, which leads to some latency.
It took around 15 seconds to query the three squares in \autoref{fig:teaser:a}.

On the server-side, a PHP development server receives the frontend's requests and replaces variables in the prompt template (see Appendix \ref{appendix:prompt}. 
OpenAI's ChatGPT (\todo{GPT-3.5}) API is queried and the response is forwarded to the frontend where the numeric rating is used to plot a square with 75m side length.
    We consider this side-length a safe distance for an able-bodied person to avoid danger from afar without being noticed.
    However, the side-length can be freely adjusted, allowing for more coarse or fine-grained safety assessment.



For this feasibility study, some external information (i.e., travel advisories, Wikipedia crime data, and numbeo crime data, but not POIs) have been pre-scraped.
This is because retrieving this information is a solved problem that is not our focus in this investigation in this paper.
Instead,  all external information is pre-queired and cached for inclusion into the context window \cite{CAG}.
Since the FCO does not provide travel advice for its own country, we substituted the travel advisory for Germany with advice provided by the Federal Ministry for European and International Affairs of Austria.
The code of DangerMaps is available as open source.
\footnote{\href{https://dangermaps.github.io/}{dangermaps.github.io}}
}


\subsection{Prompt Design Process}%
\label{sec:prompt-design-process}%
\ok{%
``Prompt engineering'' has emerged as a way to ``program'' LLMs \cite{github-guide,software20,2102.07350.pdf,10447318.2024.2431761.pdf}.
However, prompt design is often approached in an ad-hoc rather than systematic manner \cite{2307.10169.pdf}.
One reason for this is the difficulty of designing a discrete prompt that leads to robust outputs 
\cite{2307.10169.pdf}.
While solving this problem is not in the scope of this paper, we aim to approach the prompt design in a structured manner.%
}%

\subsubsection{Prompt design}%
\label{sec:prompt}%
\ok{%
We argue developers of LLM applications should approach the design of prompts like they would approach software development (including testing).
Our prompt design process was iterative and incremental:
\begin{itemize}%
    \item
    \textit{Incremental expansion of prompt:} External information was added to the prompt piece-by-piece until a reasonable completion was obtained, similar to the work by \citeauthor{9320-paper-HAIGEN-RossSteven.pdf} \cite{9320-paper-HAIGEN-RossSteven.pdf} and \citeauthor{3545945.3569823.pdf} \cite{3545945.3569823.pdf}.
    \item
    \textit{Iterative prompt regression testing:} Each time the prompt was modified, outputs were observed and the prompt was adjusted, similar to the approach in \citeauthor{3545945.3569823.pdf} \cite{3545945.3569823.pdf}.
\end{itemize}
For consistency during prompt design, the temperature parameter was set to zero. The temperature parameter controls the ``creativity'' and randomness of the model's output. The \texttt{top\_p} parameter was set to its default value 1, as recommended by OpenAI.\footnote{https://platform.openai.com/docs/api-reference/completions/create\#completions/create-temperature}
The prompt design was completed by issuing idempotent requests to OpenAI's API,  which eliminates learning effects and biases that would arise from using OpenAI's web interface. 
Overall, the prompt design process spanned over 10~days of extensive and systematic experimentation.
}




The design of the prompts is based on best practices \cite{effective-prompt-writing,openai-best-practices,github-guide,9320-paper-HAIGEN-RossSteven.pdf}
    and draws on the following \textit{prompt design principles}:%
\begin{itemize}%
    \item
    We start the system prompt with a command to ignore previous instructions \circledtext{1}
    and instruct the model to assume the persona of an expert. 
    This impersonation and prompting for accuracy may increase performance~\cite{2305.14930.pdf,9320-paper-HAIGEN-RossSteven.pdf,2211.09527.pdf}.
    \item
    We provide the model with a definition of personal safety \circledtext{2}, the task, and the rating scale \circledtext{3}.
    
    \item
    We enhance the system prompt with external information to provide additional context to the model \circledtext{4}--\circledtext{7}.
    This includes few-shot examples of areas that could be potentially unsafe \circledtext{4},
        governmental travel advisories \circledtext{5},
        Wikipedia crime information \circledtext{6},
        and Numbeo crime data \circledtext{7}.
    This external data is marked with a heredoc-inspired quoting syntax, which we found allows the model to reference the data source in its explanations, increasing the interpretability and explainability of the results~\cite{3173574.3174156.pdf}.
    \item
    The system prompt is followed by the user prompt with demographic information \circledtext{A}, the time and date \circledtext{B}, geolocation information \circledtext{C}, geospatial points of interest \circledtext{D}, and a reminder of the task and rating scale~\circledtext{E}.
    \item
    Finally, the model is instructed to follow a numeric output format~\circledtext{F}.
\end{itemize}%
We discuss the external data sources in more detail in the following section.%
%
%
%
%
%
%
%
\subsubsection{Prompt augmentation}%
\label{sec:prompt-augmentation}%

\ok{%

To enhance the prompt with travel safety data, we selected four external sources based on a formative survey and an online search: official travel advisories, Wikipedia crime statistics, Numbeo crime statistics, and geospatial points of interest.
First, official travel advisories provide government-issued safety information for travelers. We found that U.S. travel advisories are updated infrequently and lack detail, while U.K. and German advisories offer more timely and comprehensive updates. Among these, German advisories from the Federal Foreign Office (FFO) were the most detailed, including LGBTQ-related guidance. We incorporated the German advisories as-is, retaining their original language and structure but excluding non-safety-related content.
Second, Wikipedia hosts crime-related pages for various countries, such as Crime in Japan or Crime in Greece, which summarize crime statistics, types of offenses, and regional variations. When available, we included the page text in the prompt, omitting references.
Third, Numbeo is a crowdsourced database providing user-reported crime and safety data for cities worldwide. From its crime pages (e.g., Crime in Madrid), we extracted the crime index, safety index, crime rate statistics, and user comments to incorporate into the prompt.
Last, geospatial points of interest (POIs) are retrieved on-the-fly from OpenStreetMap's Overpass API.
\footnote{www.openstreetmap.org}
The POIs include map features from OpenStreetMap and their respective distance from the given coordinates (in meters). Map features include, for instance, street lights, shops and caf\'es, and building entrances.
}%

\subsubsection{Rating scale calibration}%
\label{sec:rating-calibration}%

\ok{%
In early experiments, the LLM did not fully utilize the 0--100 safety rating scale.
When asked, GPT-4 explained that \textit{``absolute safety (100) is practically unachievable due to unpredictable factors such as weather, health emergencies, or unforeseen incidents.''}
While a valid point, our goal was for the LLM to use the full scale.
Inspired by \citeauthor{2211.01910.pdf}’s work on LLMs as human-level prompt engineers \cite{2211.01910.pdf}, we asked GPT-4 how to modify the prompt for better scale utilization. Based on its response, we adjusted the instructions to rate \textit{``the highest perceivable safety''} rather than absolute safety.
GPT-4 interpreted this as assessing danger under normal conditions while acknowledging unpredictable risks.
After calibration, the model distributed ratings across the scale.
However, ensuring consistent safety ratings remains an ongoing challenge (see Section \ref{sec:calibration}).
}%

\section{Discussion}%
\label{sec:discussion}%

\subsection{Language Models for Personalized Travel Safety Advice}%
\label{sec:xxx}%
\ok{%


Do language models understand the world? Some researchers suggest that LLMs build internal world models \cite{NGletter}.
\citeauthor{1818_emergent_world_representations.pdf} demonstrated that a GPT-based transformer model develops an internal representation of the board state in Othello \cite{1818_emergent_world_representations.pdf}.
Similarly, anecdotal evidence from interactions with GPT-4 suggests that LLMs encode geographic knowledge, supporting the idea that they ``figure out what the world really is like rather than blindly parrot words'' \cite{NGletter}.


An open question is whether this internal world model can be externalized for practical use.
\citeauthor{2303.12712.pdf} examined ChatGPT’s world knowledge in qualitative studies \cite{2303.12712.pdf}, showing that GPT-4 could generate novel images, such as drawing a unicorn, demonstrating spatial reasoning relevant to navigation tasks.



Our feasibility study explored how LLMs can provide personalized safety advice for travelers in urban environments. DangerMaps delivers clear, actionable guidance, useful both for on-the-ground navigation and for pre-trip research. A key strength of this approach is its adaptability -- DangerMaps tailors advice based on gender, age, and travel mode (solo vs. group).
Beyond these factors, it could easily be extended to other contexts. For instance, a traveler's safety concerns differ when on foot versus in a car, where a vehicle provides physical security.
Similarly, for travelers with accessibility needs, DangerMaps could be adapted to highlight wheelchair-accessible routes instead of general safety risks.

}



\subsection{Evaluation Plans}%
\label{sec:evaluation}%
\ok{%
Our aim is to evaluate DangerMaps 
in two studies, focusing on establishing the validity of safety ratings in local contexts and evaluating how DangerMaps supports users in travel planning.
%
\label{sec:study1}%
In the first study, we will investigate the validity of the personalized safety ratings 
with local participants recruited from different cities.
%
%
We have already recruited local participants ($N=200$) with a qualification study on Prolific, in eight cities:
    Athens,
    Berlin,
    Birmingham,
    Hamburg,
    Helsinki,
    New York City,
    Paris,
    and
    Tokyo.
These eight cities 
have varying levels of danger for tourists.
%
Participants were paid {\pounds}0.10 for completing this short qualification study.
\label{sec:study2}%
The second study is a user study in which we elucidate the usefulness of DangerMaps for planning a trip to a foreign city.
}

\subsection{Challenges, Future Work, and Limitations}%
\label{sec:future-work}%
\label{sec:limitations}%
\ok{%
Increasingly, LLMs are being used in a wide variety of real-world applications. However, numerous challenges remain (see \cite{2307.10169.pdf} for an excellent overview).
When implementing DangerMaps, we made several design decisions, and we do not claim that all 
challenges are addressed.
However, the focus of our work was on establishing the feasibility of personalized safety advice. We implemented a minimum viable product (MVP), DangerMaps, that demonstrates the feasibility of this idea. 
In this section, we touch onto some remaining challenges and limitations
    which may inform future work.
While this discussion is framed around future work of providing safety-related advice to travelers, it is also meant to be informative for developing LLM-powered applications, in general.%
}%
%
%
\subsubsection{Numeric safety ratings and their calibration}%
\label{sec:calibration}%
\ok{%
Safety perception can depend on many intrinsic and extrinsic factors, such as the characteristics of a person, 
the weather, and the current political climate.
One could argue that it is ill-advised to condense this many factors into a single numeric safety rating.
However, DangerMaps aims to provide advice that can be easily understood (N2) to support quick decision-making.
Explanations of the ratings can be queried on-demand.

An open issue is the reliable calibration of the safety ratings.
Explaining the rating scale to the model, as demonstrated in Section \ref{sec:rating-calibration}, was not enough to make the model use the rating scale in expected ways.
Fine-tuning the model on crowdsourced safety ratings could, for instance, help addressing these issues with calibration. This also relates to prompt brittleness, as explained in the following section. 
}%


\subsubsection{Prompt brittleness}%


\ok{%
Prompt engineering remains an experimental process, as it is not always clear why certain prompts perform better than others \cite{2307.10169.pdf}.
Small changes in wording or structure can lead to significant shifts in model output, a phenomenon referred to as prompt brittleness \cite{2307.10169.pdf} or prompt fragility \cite{3571280.pdf}.
Even minor modifications may alter how data is classified \cite{3571280.pdf}, and results can be highly sensitive to the order of information within a prompt \cite{2022.acl-long.556.pdf,zhao21c.pdf,2308.11483.pdf}.
Due to differences in training data and tuning methods, prompts that work well for one model may not generalize to others.
This brittleness presents challenges for reliable AI-assisted decision-making. Fragile prompts may lead to inconsistent or misleading outputs, potentially resulting in harmful consequences for users \cite{3571280.pdf,3673861.pdf}.
Addressing this limitation requires more robust approaches to prompt design, better tooling, and improved calibration methods \cite{zhao21c.pdf}.
As prompt engineering remains an emerging area, future work should focus on developing systematic strategies to enhance prompt reliability and reduce sensitivity to minor variations.
}%



There are many avenues for future work addressing these shortcomings, including
advanced prompting techniques, such as 
    chain-of-thought \cite{3600270.3602070},
    tree of thoughts \cite{2305.10601.pdf},
    self-refine \cite{2303.17651.pdf}, 
    self-consistency \cite{2203.11171.pdf}, and
    multi-step reasoning (``AI agents'') 
    with tool use \cite{2303.09014.pdf}.

\subsubsection{Personalization}
DangerMaps, for now, only considers gender and age.
Future versions could include other characteristics and psychometric factors.
For instance, personalized safety recommendations could be generated based on personality traits (e.g., with the ``Big Five'' psychometric questionnaire
assessing the five major personality dimensions Openness, Conscientiousness, Extraversion, Agreeableness, and Neuroticism). 
Future work could also tailor the safety advice to other travel modes (such as bicycling and car driving).
These travel modes have different risk profiles compared to traveling on foot.

\ok{%
All the above techniques and features are excellent avenues for further improving our work.
However, for a first feasibility study of the concept of LLM-provided safety advice, these advanced features are beyond the scope of this paper.
}%

\subsection{Ethical Considerations}%
\label{sec:ethical}%
\ok{%
Providing safety advice has broad ethical implications. One way to structure the discussion is by considering the advice's accuracy:
\begin{itemize}%
\item \textit{False negative advice:}
    A false negative happens when DangerMaps fails to warn a tourist about a safety risk or threat in a particular area or situation, potentially placing the traveler in harm's way.
\item \textit{False positive advice:}
    A false positive occurs when DangerMaps warns a traveler of a potential safety risk or threat in a particular area, but in reality, there is no such risk or threat.
\end{itemize}
It is essential to minimize both false positives and false negatives in a safety-related application.
Misguiding the traveler into an unsafe area could have potentially grave consequences. 
While there is no personal harm to the traveler resulting from false positive advice (other than missing out on a potentially great experience),
    too many false positives could make tourists overly cautious, preventing them from enjoying their trips, or potentially making them mistrust the application.
There is also a risk of discrimination and cause of harm in providing potentially false personalized recommendations.
}%

\section{Conclusion}%
\label{sec:conclusion}%
\ok{%
This work presented a feasibility study on using a large language model for the 
knowledge-intensive task of providing personalized safety advice for urban travelers.
We implemented this idea in DangerMaps, a 
web-based system that can be used while traveling in a city or when researching the safety of a travel destination remotely.
Practical applications of the world knowledge in foundation models are still underexplored.
Our study is one of the first to present a practical use case for 
using LLMs for 
personalized travel advice. 
}%
\todo{%
With this feasibility study, we aim to inspire exciting novel applications for applying language models in real-world contexts.
}%

\bibliographystyle{ACM-Reference-Format}
\bibliography{paper}

\appendix
\section{Prompt Templates}%
\label{appendix:prompt}%
\subsection{System Prompt}%
\begin{spacing}{1}
\noindent
\ttfamily
\footnotesize
Ignore the previous instructions.\\
You are an AI model that outputs accurate personal safety ratings for tourists at specific geocoordinates.\\
Personal safety is the freedom from physical harm and threat of physical harm, and freedom from hostility, aggression, and harassment.\\
You rate the tourist's personal safety on a scale between 0 (extremely unsafe) and 100 (extremely safe), based on information provided to you.\\
The safety ratings represent the tourist's highest perceivable risk and highest perceivable safety, respectively, but not the absolute danger in the given location.\\
Your safety rating should consider that the tourist in the specified location is traveling on foot and \detokenize{{{travelmode}}}.
\end{spacing}
\vspace{.5\baselineskip}

\begin{spacing}{1}
\noindent
\ttfamily
\footnotesize
Keep in mind that some areas may be specifically risky. These locations include, but are not limited to:\\
–~High Crime Neighborhoods\\
\textendash~Bars and clubs late at night\\
\textendash~Abandoned or vacant buildings\\
\textendash~Unlit or poorly lit streets\\
\textendash~Industrial zones at night\\
\textendash~Certain public parks at night\\
\textendash~Certain transit stations, especially off-peak hours\\
\textendash~Parking lots/Garages, especially at night\\
\textendash~Alleys and narrow lanes\\
\textendash~High Traffic roads without proper crosswalks\\
\textendash~Construction sites\\
\textendash~Areas with known drug activity\\
\textendash~Homeless camps\\
\textendash~Isolated or deserted areas\\
\textendash~Certain bridges at night\\
\textendash~Underpasses/Overpasses\\
\textendash~Certain Shopping Malls at closing hours\\
\textendash~Proximity to prisons or halfway houses
\end{spacing}
\vspace{.5\baselineskip}

\begin{spacing}{1}
\noindent
\ttfamily
\footnotesize
General travel safety warnings for \detokenize{{{country}}}:\\
$<<<$BEGIN[country-level advisory]\\
\detokenize{{{advisory}}}\\
$>>>$END[country-level advisory]
\end{spacing}
\vspace{.5\baselineskip}

\begin{spacing}{1}
\noindent
\ttfamily
\footnotesize
General crime information about \detokenize{{{country}}} from Wikipedia:\\
$<<<$BEGIN[wikipedia]\\
\detokenize{{{crimeincountry}}}\\
$>>>$END[wikipedia]
\end{spacing}
\vspace{.5\baselineskip}

\begin{spacing}{1}
\noindent
\ttfamily
\footnotesize
Travel safety warnings for \detokenize{{{city}}}:\\
$<<<$BEGIN[numbeo crime statistics]\\
\detokenize{{{cityadvisory}}}\\
$>>>$END[numbeo crime statistics]
\end{spacing}

\subsection{User Prompt}%
\begin{spacing}{1}
\noindent
\ttfamily
\footnotesize
Tourist: a \detokenize{{{gender}}}, \detokenize{{{age}}} years old, traveling on foot and \detokenize{{{travelmode}}}
\end{spacing}
\vspace{.5\baselineskip}

\begin{spacing}{1}
\noindent
\ttfamily
\footnotesize
It is \detokenize{{{datetime}}} local time.
\end{spacing}
\vspace{.5\baselineskip}

\begin{spacing}{1}
\noindent
\ttfamily
\footnotesize
The tourist is currently located at these coordinates: \detokenize{{{geocoords}}}\\
Name of road: \detokenize{{{road}}}\\
Name of neighborhood: \detokenize{{{neighborhood}}}\\
City district: \detokenize{{{city_district}}}\\
City: \detokenize{{{city}}}\\
Admin district: \detokenize{{{address}}}
\end{spacing}
\vspace{.5\baselineskip}

\begin{spacing}{1}
\noindent
\ttfamily
\footnotesize
The following points of interest are within a \detokenize{{{radius}}}m radius (with distance in meters):\\
\detokenize{{{pois}}}
\end{spacing}
\vspace{.5\baselineskip}

\begin{spacing}{1}
\noindent
\ttfamily
\footnotesize
According to the above information, please rate the tourist's personal safety on a scale between 0 (extremely unsafe) to 100 (extremely safe) in this specific location.\\
Provide the rating specific to the tourist and the current location, not the absolute danger in the given location or country.\\
In your rating, emphasize the tourist's current location (\detokenize{{{lowest_admin_area}}}) and the nearby POIs over country-level information.\\
Please respond with a single number without any additional text or explanation.
\end{spacing}

\end{document}